\documentclass[fleqn,twoside]{article}
\usepackage{espcrc2,amsmath}

\usepackage{graphicx}

\newcommand{\cO}{{\cal O}}

\newcommand{\as}{\alpha_s}
\newcommand{\wh}{\widehat}
\newcommand{\nn}{\nonumber}

\newcommand{\eqn}[1]{(\ref{#1})}

\newcommand{\gev}{\mbox{\rm GeV}}
\newcommand{\tvs}{\vbox{\vskip 4mm}}

\newcommand{\sfrac}[2]{\mbox{$\frac{#1}{#2}$}}

\title{$\alpha_s$ and the $\tau$ hadronic width}

\author{Matthias Jamin\address{Instituci\'o Catalana de Recerca i Estudis
        Avan\c{c}ats (ICREA), IFAE, Theoretical Physics Group, UAB,
        E-08193 Bellaterra, Barcelona, Spain}}

\begin{document}

\begin{abstract}
Different choices exist for the renormalisation group resummation in the
determination of $\as$ from hadronic $\tau$ decays: namely fixed-order (FOPT)
and contour-improved perturbation theory (CIPT). The two approaches lead to
systematic differences in the resulting $\as$. On the basis of a model for
higher-order terms in the perturbative series, which incorporates well-known
structure from renormalons, it is found that while CIPT is unable to account
for the fully resummed series, FOPT smoothly approaches the Borel sum.
Employing the model to determine $\as$ yields $\as(M_\tau)=0.316 \pm 0.006$,
which after evolution leads to $\as(M_Z) = 0.1180 \pm 0.0008$.
\vspace{1pc}
\end{abstract}

\maketitle

\section{INTRODUCTION}

Due to its particular mass of $M_\tau\approx 1.8\,\gev$, the $\tau$ lepton
constitutes an excellent system to study QCD at low energies, as in about 65\%
of all cases it decays into hadrons, while the QCD description remains
largely perturbative. In the seminal work \cite{bnp92}, the framework for
a precise determination of the QCD coupling $\as$ from the total $\tau$
hadronic width
\begin{equation}
\label{Rtauex}
R_\tau \,\equiv\, \frac{\Gamma[\tau^- \to {\rm hadrons} \, \nu_\tau (\gamma)]}
{\Gamma[\tau^- \to e^- \overline \nu_e \nu_\tau (\gamma)]} \,=\,
3.640(10) \,,
\end{equation}
was developed, while afterwards invariant mass distributions were incorporated
into the analysis as well \cite{aleph05,dhz05}. The most recent study of the
ALEPH spectral function data on the basis of the final full LEP data set yielded
$\as(M_\tau)=0.344\pm 0.005_{\rm exp}\pm 0.007_{\rm th}$ \cite{ddhmz08}, which
after evolution to the $Z$-boson mass scale results in
$\as(M_Z)=0.1212(11)$. The dominant quantifiable theory uncertainty resides
in as yet uncalculated higher-order QCD corrections and improvements of the
perturbative series through renormalisation group (RG) methods.

Most suitable for the $\as$ determination is the $\tau$ decay rate into light
$u$ and $d$ quarks $R_{\tau,V/A}$ via a vector or axialvector current, since
in this case power corrections are especially suppressed. Theoretically,
$R_{\tau,V/A}$ takes the form \cite{bnp92}
\begin{eqnarray}
\label{RtauVA}
R_{\tau,V/A} &\!\!=\!\!& \frac{N_c}{2}\,S_{\rm EW}\,|V_{ud}|^2\,\Big[\,
1 + \delta^{(0)} \nn \\
&& +\,\delta_{\rm EW}' + \sum\limits_{D\geq 2} \delta_{ud,V/A}^{(D)} \,\Big]\,,
\end{eqnarray}
where $S_{\rm EW}=1.0198(6)$ \cite{ms88} and $\delta_{\rm EW}'=0.0010(10)$
\cite{bl90} are electroweak corrections, $\delta^{(0)}$ comprises the
perturbative QCD correction, and the $\delta_{ud,V/A}^{(D)}$ denote quark
mass and higher $D$-dimensional operator corrections which arise in the
framework of the operator product expansion (OPE).

When computing the total $\tau$ hadronic width a phase-space integral over the
physical spectrum has to be calculated, which by analyticity can be related to
a contour integral over QCD correlators in the complex $s$-plane, where $s$ is
the invariant mass of the final state hadronic system. As we also intend to
RG improve the perturbative series, the questions arises if the RG resummation
should be performed before or after evaluating the contour integral? If the
``true'' all-order result were available, both treatments should agree, but to
any finite order in perturbation theory, significant differences may arise.

The approach of first RG-improving the correlators and then performing the
contour integral was introduced in \cite{piv91,dp92} and termed contour-improved
perturbation theory (CIPT), as unarguably large running effects of
$\as(\sqrt{s})$ along the contour are resummed. However, it is known that the
QCD series is divergent, being asymptotic at best. Hence, also the explicit
expansion coefficients of the perturbative series are bound to become large.
If cancellations between explicit coefficients and running effects occur, just
resumming the running effects may not lead to a good approximation, but it
could be better to perform a consistent expansion in powers of $\as(M_\tau)$,
being called fixed-order perturbation theory (FOPT).

While it can be demonstrated that, as expected, CIPT is a good approximation,
and CIPT as well as FOPT are compatible, when the running effects dominate
\cite{jam05}, in the large-$\beta_0$ approximation, in which $\delta^{(0)}$ is
exactly calculable, FOPT gives a better approximation to the true result and
CIPT is not seen to be compatible with it \cite{bbb95,ben98}. Thus it is not a
priori clear which behaviour prevails in real QCD. It should be obvious that
this is a question about perturbative orders beyond the presently known
${\cal O}(\as^4)$ term \cite{bck08}.

To further investigate the difference between CIPT and FOPT, the contribution
of higher perturbative orders has to be modelled. This then allows to
investigate the following questions:
\begin{itemize}

\item[i)] Are FO and CI perturbation theory found to be compatible, once terms
beyond the currently known perturbative coefficients of the series for
$\delta^{(0)}$ are included?

\item[ii)] How do FO and CI perturbation theory at a particular order compare
to the true result for $\delta^{(0)}$, and is the closest approach
to the true  result related to the minimal terms in the respective series?

\item[iii)] And finally, which of the two methods, FOPT or CIPT provides the
closer approach to the true value at order ${\cal O}(\as^4)$, and in
general?

\end{itemize}
The working assumption in all this is that the ``true'' result is approximated
with reasonable accuracy by the Borel sum of the model series under
consideration, since the power corrections to $R_\tau$ are known to be small.
The model employed below will be constructed such as to incorporate the general
structure of the Adler function in the Borel plane, dictated by the OPE and
the RG.

\begin{boldmath}
\section{PERTURBATIVE CORRECTION $\delta^{(0)}$}
\end{boldmath}

In the following, we shall only be concerned with the purely perturbative
correction $\delta^{(0)}$ which gives the dominant contribution to
$R_{\tau,V/A}$. In FOPT it takes the general form
\begin{equation}
\label{del0FO}
\delta^{(0)}_{\rm FO} \,=\, \sum\limits_{n=1}^\infty a(M_\tau^2)^n
\sum\limits_{k=1}^{n} k\,c_{n,k}\,J_{k-1} \,,
\end{equation}
where $a(\mu^2)\equiv a_\mu\equiv\as(\mu)/\pi$, and $c_{n,k}$ are the
coefficients which appear in the perturbative expansion of the vector
correlation function,
\begin{equation}
\label{Pis}
\Pi_V(s) \,=\, -\,\frac{N_c}{12\pi^2} \sum\limits_{n=0}^\infty a_\mu^n
\sum\limits_{k=0}^{n+1} c_{n,k} \ln^k\!\left(\frac{-s}{\mu^2}\right) .
\end{equation}
At each perturbative order, the coefficients $c_{n,1}$ can be considered
independent, while all other $c_{n,k}$ with $k\geq 2$ are calculable from the
RG equation. Further details can for example be found in ref.~\cite{bj08}.
Finally, the $J_l$ are contour integrals which are defined by
\begin{equation}
\label{Jl}
J_l \,\equiv\, \frac{1}{2\pi i} \!\!\oint\limits_{|x|=1} \!\!
\frac{dx}{x}\, (1-x)^3\,(1+x) \ln^l(-x) \,.
\end{equation}
The first three which are required up to $\cO(\as^3)$ take the numerical values
\begin{equation}
\label{J0to2}
J_0 \,=\, 1 \,, \quad
J_1 \,=\, -\,\sfrac{19}{12} \,, \quad
J_2 \,=\, \sfrac{265}{72} - \sfrac{1}{3}\,\pi^2 \,.
\end{equation}

At order $\as^n$ FOPT contains unsummed logarithms of order 
$\ln^l(-x)\sim \pi^l$ with $l<n$ related to the contour integrals $J_l$.
CIPT sums these logarithms, which yields
\begin{equation}
\label{del0CI}
\delta^{(0)}_{\rm CI} \,=\, \sum\limits_{n=1}^\infty c_{n,1}\,
J_n^a(M_\tau^2)
\end{equation}
in terms of the contour integrals $J_n^a(M_\tau^2)$ over the running coupling,
defined as:
\begin{equation}
\label{Jna}
J_n^a(M_\tau^2) \,\equiv\, \frac{1}{2\pi i} \!\!\oint\limits_{|x|=1}\!\!
\frac{dx}{x}\,(1-x)^3\,(1+x)\,a^n(-M_\tau^2 x) \,.
\end{equation}
In contrast to FOPT, for CIPT each order $n$ just depends on the corresponding
coefficient $c_{n,1}$. Thus, all contributions proportional to the coefficient
$c_{n,1}$ which in FOPT appear at all perturbative orders equal or greater than
$n$ are resummed into a single term.

Numerically, the two approaches lead to significant differences. Using
$\as(M_\tau)=0.34$ in eqs.~\eqn{del0FO} and \eqn{del0CI}, one finds
\begin{eqnarray}
\label{del0FOn}
\delta^{(0)}_{\rm FO} &\!\!=\!\!& 0.2200 \;(0.2288) \,, \\
\tvs
\label{del0CIn}
\delta^{(0)}_{\rm CI} &\!\!=\!\!& 0.1984 \;(0.2021) \,,
\end{eqnarray}
where the first number in both cases employs the known coefficients up to
$\cO(\as^4)$ \cite{bck08} and the numbers in brackets include an estimate of
the $\cO(\as^5)$ term with $c_{5,1}\approx 283$ \cite{bj08}. Inspecting the
individual contributions from each order, up to $\cO(\as^5)$ the CIPT series
appears to be better convergent. However, around the seventh order, the
contour integrals $J_n^a(M_\tau^2)$ change sign and thus at this order the
contributions are bound to become small. Therefore, the faster approach to the
minimal term does not necessarily imply that CIPT gives the closer approach to
the true result for the resummed series.

\vskip 6mm 
\section{A PHYSICAL MODEL}

To clarify whether FOPT or CIPT results in a better approximation to
$\delta^{(0)}$, one needs to construct a physically motivated model for its
series. The corresponding model will be based on the Borel transform of the
Adler function $D_V(s)$:
\begin{equation}
\label{DVs}
D_V(s) \,\equiv\, -\,s\,\frac{d}{ds}\,\Pi_V(s) \,\equiv\,
\frac{N_c}{12\pi^2}\,\big[ 1+\wh D(s) \big] \,.
\end{equation}
In the following discussion it is slightly more convenient to utilise the
related function $\wh D(s)$. Its Borel transform $B[\wh D](t)$ is defined by
the relation
\begin{equation}
\label{Dalpha}
\wh D(\alpha) \,\equiv\, \int\limits_0^\infty dt\,{\rm e}^{-t/\alpha}\,
B[\wh D](t)\,.
\end{equation}
The integral $\wh D(\alpha)$, if it exists, gives the Borel sum of the original
divergent series. It was found that the Borel-transformed Adler function
$B[\wh D](t)$ obtains infrared (IR) and ultraviolet (UV) renormalon poles at
positive and negative integer values of the variable $u\equiv 9t/(4\pi)$,
respectively \cite{ben93,bro93}. (With the exception of $u=1$.)

Apart from very low orders, where a dominance of renormalon poles close to
$u=0$ has not yet set in, intermediate orders should be dominated by the
leading IR renormalon poles, while the leading UV renormalon, being closest
to $u=0$, dictates the large-order behaviour of the perturbative expansion.
Assuming that only the first two orders are not yet dominated by the lowest
IR renormalons, one is led to the ansatz
\begin{eqnarray}
\label{BRu}
\hspace{-2mm} B[\wh D](u) &\!\!=\!\!& B[\wh D_1^{\rm UV}](u) +
B[\wh D_2^{\rm IR}](u) + \nn \\
\tvs
&& B[\wh D_3^{\rm IR}](u) + d_0^{\rm PO} + d_1^{\rm PO} u \,,
\end{eqnarray}
which includes one UV renormalon at $u=-1$, the two leading IR renormalons at
$u=2$ and $u=3$, as well as polynomial terms for the two lowest perturbative
orders. Explicit expressions for the UV and IR renormalon pole terms
$B[\wh D_p^{\rm UV}](u)$ and $B[\wh D_p^{\rm IR}](u)$ can be found in section~5
of ref.~\cite{bj08}.

Apart from the residues $d_p^{\rm UV}$ and $d_p^{\rm IR}$, the full structure
of the renormalon pole terms is dictated by the OPE and the RG. Therefore, the
model~\eqn{BRu} depends on five parameters, the three residua $d_1^{\rm UV}$,
$d_2^{\rm IR}$ and $d_3^{\rm IR}$, as well as the two polynomial parameters
$d_0^{\rm PO}$ and $d_1^{\rm PO}$. These parameters can be fixed by matching
to the perturbative expansion of $\wh D(s)$ up to $\cO(\as^5)$. Thereby one
also makes use of the estimate for $c_{5,1}$. The parameters of the model
\eqn{BRu} are then found to be:
\begin{equation}
\label{dUVIRa}
d_1^{\rm UV} =\, -\,1.56\cdot 10^{-2} \,,\;
d_2^{\rm IR} =\,    3.16  \,,\;
d_3^{\rm IR} =\, -\,13.5 \,, \nn \\[-1mm]
\end{equation}
\begin{equation}
\label{dUVIRb}
d_0^{\rm PO} =\,    0.781 \,,\;
d_1^{\rm PO} =\,    7.66\cdot 10^{-3} \,.
\end{equation}
The fact that the parameter $d_1^{\rm PO}$ turns out to be small implies that
the coefficient $c_{2,1}$ is already reasonably well described by the
renormalon pole contribution, although it was not used to fix the residua.
Therefore, one could set $d_1^{\rm PO}=0$ and actually work with a model which
only has four parameters. The predicted value $c_{5,1}=280$ in this model turns
out very close to the estimate, which can be viewed as one test of the stability
of the model.

\begin{figure}[thb]
\hspace{-3mm}
\includegraphics[angle=0, width=7.8cm]{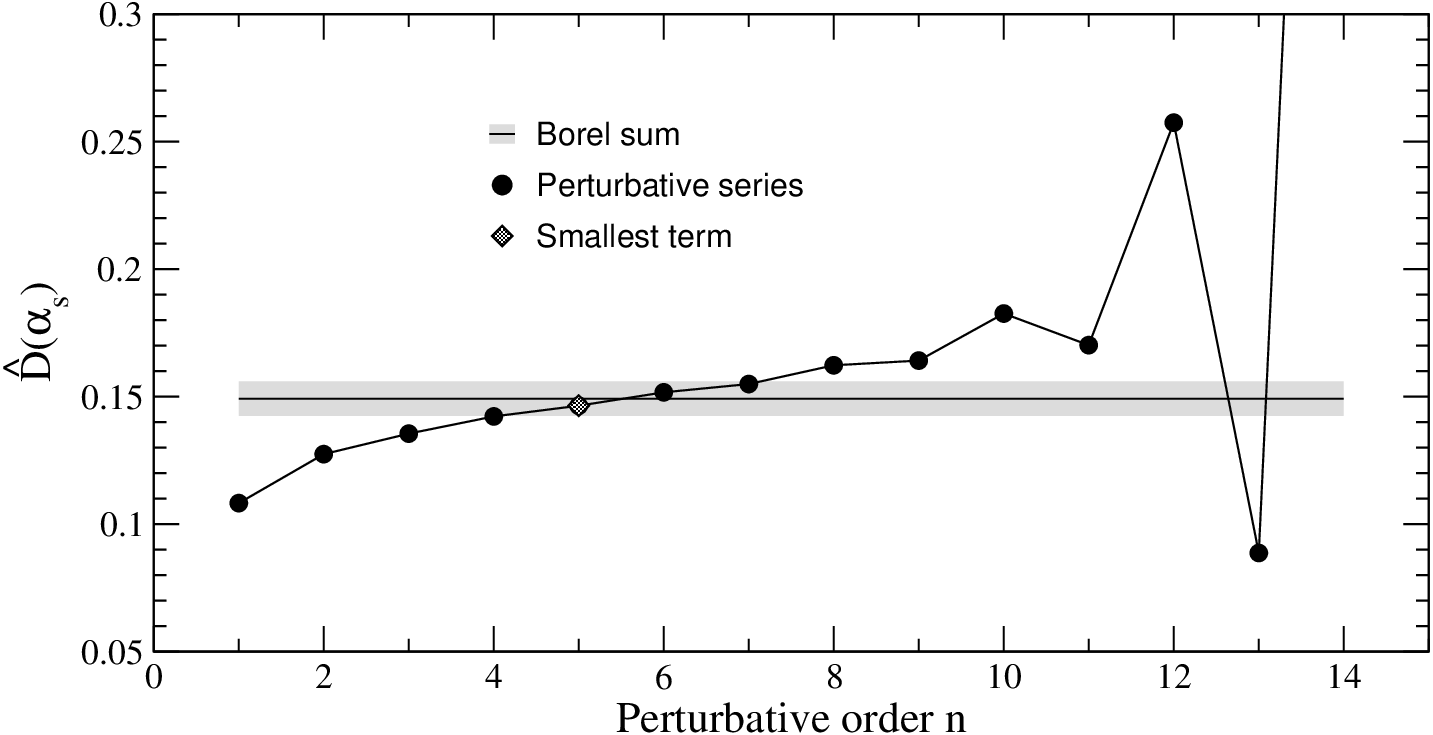}
\vspace{-1cm}
\caption{Results for $\wh D(M_\tau^2)$ (full circles) at $\as(M_\tau)=0.34$,
employing the model \eqn{BRu}, as a function of the order $n$ up to which the
terms in the perturbative series have been summed. The straight line represents
the result for the Borel sum of the series.\label{figadler}}
\end{figure}

A graphical account of the model \eqn{BRu} for the (reduced) Adler function
$\wh D(M_\tau^2)$ is displayed in figure~\ref{figadler}. The full circles
denote the partial sums of the perturbative series up to order $n$. The minimal
term of the series at the 5th order is marked by a grey diamond. The
perturbative results are compared with the Borel sum of the model (straight
line).\footnote{Also shown as the shaded band is an estimate of the uncertainty
inferred from the complex ambiguity which arises while defining the Borel
integral over the IR renormalon poles. For details see appendix~A of
ref.~\cite{bj08}.} Generally, figure~\ref{figadler} shows that the model is
well-behaved: the series goes through a number of small terms such that the
truncated series nicely agrees with its Borel sum, before the sign-alternating
asymptotic behaviour takes over around $n=10$.

\begin{figure}[thb]
\hspace{-3mm}
\includegraphics[angle=0, width=7.8cm]{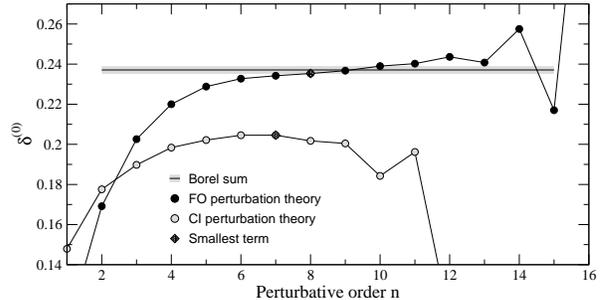}
\vspace{-1cm}
\caption{Results for $\delta^{(0)}_{\rm FO}$ (full circles) and
$\delta^{(0)}_{\rm CI}$ (grey circles) at $\as(M_\tau)=0.34$, employing the
model \eqn{BRu}, as a function of the order $n$ up to which the terms in the
perturbative series have been summed. The straight line represents the result
for the Borel sum of the series.\label{fig6}}
\end{figure}

The implications of the model \eqn{BRu} for $\delta^{(0)}$ in FOPT and CIPT is
graphically represented in figure~\ref{fig6}. The full circles denote the
result for $\delta^{(0)}_{\rm FO}$ and the grey circles the one for
$\delta^{(0)}_{\rm CI}$, as a function of the order $n$ up to which the
perturbative series has been summed. The straight line corresponds to the
principal value Borel sum of the series, $\delta^{(0)}_{\rm BS}=0.2371$, and
the shaded band provides an error estimate based on the imaginary part divided
by $\pi$. The order at which the series have their smallest terms is indicated
by the grey diamonds. As is obvious from figure~\ref{fig6}, like for the
Adler function itself, FOPT displays the behaviour expected from an asymptotic
series: the terms decrease up to a certain order around which the closest
approach to the resummed result is found, and for even higher orders, the
divergent large-order behaviour of the series sets in. For CIPT, on the other
hand, the asymptotic behaviour sets in earlier, and the series is never able
to come close to the Borel sum.

The finding that in the model \eqn{BRu} CIPT misses the full Borel sum can be
traced back to the fact that in CIPT the running effects along the complex
contour are resummed to all orders, while explicit contributions of the
$c_{n,1}$ at a certain order are dropped. However, being an asymptotic series,
also the Adler function coefficients $c_{n,1}$ become large, and cancellations
between the explicit contributions and the running effects take place. As was
shown in section~4 of \cite{bj08}, in the large-$\beta_0$ approximation, for
$R_\tau$ the leading IR renormalon cancels completely, and also the large-order
divergence of the series is softened. Even though in real QCD the leading IR
does not anymore cancel completely, for $R_\tau$ it is still suppressed by a
factor $1/n^2$, and furthermore a sign-alternating UV renormalon component does
not yet show up in the known coefficients. Thus, the cancellations between
running effects and explicit coefficients are also expected to prevail in full
QCD.

The deficiency of CIPT to approach the Borel sum of the series, which leads to
the marked differences of $\delta^{(0)}_{\rm FO}$ and $\delta^{(0)}_{\rm CI}$,
can also be observed when the Adler function is inspected along the complex
circle. This is discussed in detail in appendix~B of ref.~\cite{bj08}. While
FOPT converges to the Borel sum on the full circle (though rather badly close
to the Minkowskian axis), in some regions of the circle CIPT largely differs
from the resummed result, again due to the missed cancellations.

As the behaviour of CIPT versus FOPT hinges on the contribution of the leading
IR renormalon at $u=2$, in principal also models can be constructed for which
CIPT provides a good account of the Borel sum. These would generally be models
where $d_2^{\rm IR}$ is much smaller than the value quoted in eq.~\eqn{dUVIRb}.
While such models can at present not be excluded, the pattern of the individual
contributions appears more unnatural than in the model \eqn{BRu}: the known
$c_{n,1}$ can only be reproduced when one allows for large cancellations
between the individual terms. Thus, the behaviour generally expected from the
presence of renormalon poles, namely dominance of leading IR poles at
intermediate orders, would be lost.

\begin{boldmath}
\section{DETERMINATION OF $\as$}
\end{boldmath}

The starting point for a determination of $\as$ from hadronic $\tau$ decays
is eq.~\eqn{RtauVA} for the decay rate of the $\tau$ lepton into light $u$
and $d$ quarks. The analysis will be based on FOPT together with the ansatz
\eqn{BRu} for higher-order terms discussed in the last section. Due to the
observation that CIPT is not able to approach the resummed series, it will
not be employed below.

The first step of the $\as$ analysis consists in estimating the values of the
power corrections $\delta_{ud,V+A}^{(D)}$, which arise from higher-dimensional
operators in the framework of the OPE. Given these estimates and experimental
data, a phenomenological value of $\delta^{(0)}$ can be calculated using
eq.~\eqn{RtauVA}. This allows to determine the value of $\as(M_\tau)$ by
requiring that the theoretical value $\delta^{(0)}_{\rm FO}$ matches the
phenomenological value $\delta^{(0)}_{\rm phen}$. Errors are estimated by
varying all parameters within their uncertainties.

In view of the smallness of the light quark masses $m_u$ and $m_d$, as well as
the suppression of the dimension-4 contribution in $R_\tau$, the dominant power
correction arises from the six dimensional 4-quark condensates. As the number
of contributing operators is too large to treat the 4-quark condensates
individually, conventionally the so-called vacuum-saturation approximation
(VSA) \cite{svz79} is employed. Then the corresponding contribution takes the
form
\begin{eqnarray}
\label{del6}
\delta_{\langle\bar qq\bar qq\rangle,V+A}^{(6)} &\!\!=\!\!& -\,\frac{512}{27}
\,\pi^3 \as\,\frac{\rho\langle\bar qq\rangle^2}{M_\tau^6} \nn \\
&\!\!=\!\!& ( -\, 4.8 \pm 2.9 )\cdot 10^{-3} \,,
\end{eqnarray}
where for the numerical estimate the required quark condensate is taken from
the GMOR relation \cite{gmor68,jam02}, and $\rho=2\pm 1$ was assumed to take
into account violations of the VSA. This choice includes most estimates of the
four-quark condensate present in the literature.

To complete the estimate of power corrections to $R_\tau$, the longitudinal
contribution which arises from the pseudoscalar correlator still has to be
included.\footnote{The scalar correlator, being proportional to $(m_u-m_d)^2$,
is completely negligible.} Because the perturbative series for this correlator
does not converge very well, the approach of refs.~\cite{gjpps03,gjpps04} will
be followed. The main idea is to replace the QCD expression for the pseudoscalar
correlator by a phenomenological representation. The dominant contribution to
the pseudoscalar spectral function stems from the well-known pion pole, giving
\begin{equation}
\label{delPSpi}
\delta_{ud,S+P}^\pi \,=\, -\,16\pi^2\,\frac{f_\pi^2 M_\pi^2}{M_\tau^4}
\Biggl(1-\frac{M_\pi^2}{M_\tau^2}\Biggr)^{\!2} \,,
\end{equation}
plus small corrections from higher-excited pionic resonances. Repeating the
analysis of ref.~\cite{gjpps03} and updating the input parameters, one finds
\begin{equation}
\label{delspin0}
\delta_{ud,S+P} \,=\, ( -\,2.64 \pm 0.05 )\cdot 10^{-3} \,.
\end{equation}
Collecting all contributions, and adding the errors in quadrature, one arrives
at the total estimate of all power corrections:
\begin{equation}
\label{delph}
\delta_{\rm PC} \,=\, ( -\, 7.1 \pm 3.1 )\cdot 10^{-3} \,.
\end{equation}
The value \eqn{delph} is consistent with the most recent fit to the $\tau$
spectral functions performed in ref.~\cite{ddhmz08}.

As a matter of principle, the OPE of correlation functions in the complex
$s$-plane could be inflicted with so-called ``duality violations'' \cite{shi00}.
These arise from the contour integral close to the physical region which even
though suppressed in $R_\tau$ could be sizeable \cite{cgp08}. Nevertheless,
before a possible additional duality violating contribution can be extracted
consistently from a combined fit to spectral moments, it shall be omitted.

Employing the value $R_{\tau,V+A}=3.479\pm 0.011$, which results from
eq.~\eqn{Rtauex} in conjunction with $R_{\tau,S}= 0.1615\pm 0.0040$
\cite{ddhmz08}, as well as $|V_{ud}|=0.97418\pm 0.00026$ \cite{th07}, from
eqs.~\eqn{RtauVA} and \eqn{delph} the phenomenological value for $\delta^{(0)}$
can be derived:
\begin{equation}
\label{de0ph}
\delta^{(0)}_{\rm phen} \,=\, 0.2042 \pm 0.0050 \,.
\end{equation}
The dominant experimental uncertainty in \eqn{de0ph} is due to $R_{\tau,V+A}$
and the theoretical one to the dimension-6 condensate. The final step in the
extraction of $\as(M_\tau)$ now consists in finding the values of $\as$ for
which $\delta^{(0)}_{\rm phen}$ matches the theoretical prediction, which
yields \cite{bj08}
\begin{equation}
\label{astau}
\as(M_\tau) \,=\, 0.3156 \pm 0.0030_{\rm exp} \pm 0.0051_{\rm th} \,.
\end{equation}
Evolving this result to the $Z$-boson mass scale finally leads to
\begin{equation}
\label{asMz}
\as(M_Z) \,=\, 0.11795 \pm 0.00076 \,,
\end{equation}
in perfect agreement with the world average \cite{pdg06}.

\vskip 4mm 
\section*{Acknowledgements}
The author would like to thank Martin~Beneke for a most enjoyable collaboration.
This work has been supported in parts by EU Contract No. MRTN-CT-2006-035482
(FLAVIAnet), by CICYT-FEDER-FPA2005-02211, and by the Spanish Consolider-Ingenio
2010 Programme CPAN (CSD2007-00042).

\providecommand{\href}[2]{#2}\begingroup\raggedright\endgroup

\vfill

\end{document}